\documentclass[epj]{webofc}
\usepackage[utf8]{inputenc}
\usepackage[varg]{txfonts}   
\usepackage{booktabs}
\usepackage{xcolor}
\definecolor{darkred}{rgb}{0.4,0.0,0.0}
\definecolor{darkgreen}{rgb}{0.0,0.4,0.0}
\definecolor{darkblue}{rgb}{0.0,0.0,0.4}
\usepackage[bookmarks,linktocpage,colorlinks,
    linkcolor = darkred,
    urlcolor  = darkblue,
    citecolor = darkgreen]{hyperref}
\usepackage{subfigure}
\wocname{EPJ Web of Conferences}
\woctitle{Lattice2017}
\begin{document}
%
\selectlanguage{english}
\title{%
Duals of U(N) LGT with staggered fermions
}
\author{%
\firstname{Oleg} \lastname{Borisenko}\inst{1}\fnsep\thanks{Speaker, \email{oleg@bitp.kiev.ua}} \and
\firstname{Volodymyr} \lastname{Chelnokov}\inst{1} \and 
\firstname{Sergey}  \lastname{Voloshyn}\inst{1}
}
\institute{%
Institute for Theoretical Physics of the National Academy of Sciences of Ukraine, \\ 
14-b Metrolohichna str., 03143 Kiev, Ukraine 
}
\abstract{%
Various approaches to construction of dual formulations of non-abelian 
lattice gauge theories are reviewed. In the case of U(N) LGT we use 
a theory of the Weingarten functions to construct a dual formulation. 
In particular, the dual representations are constructed 1) for pure gauge 
models in all dimensions, 2) in the strong coupling limit for the models 
with arbitrary number of flavours and 3) for two-dimensional U(N) QCD with 
staggered fermions. Applications related to the finite temperature/density 
QCD are discussed.
}
\maketitle
\section{Duals of lattice spin and gauge models}\label{intro} 

Dual representations proved to be a very useful concept in the context 
of abelian spin and gauge models. The application of dual representations 
ranges from the determination of the critical points in the self-dual abelian 
models to the proof of the confinement in the three-dimensional $U(1)$ lattice 
gauge theory (LGT) \cite{dualu1,mack} and the numerical study of the $U(1)$ LGT 
both at zero \cite{zachdiss} and at finite temperature \cite{dual_u1_dec}.
For abelian models the dual transformations are well-defined and described 
in many reviews and text books \cite{savit}. The status of the dual representations 
of non-abelian models is very different. During decades several different approaches 
have been attempted to construct dual representations. 
\begin{itemize} 
\item 
Dual representations based on the plaquette formulation \cite{halpern,plrepr,plaq_gauge}. 
Dual variables are introduced as variables conjugate to local Bianchi identities \cite{halpern,math_probl,conrady}. 
The dual model appears to be non-local due to the presence of connectors in the Bianchi identities for 
gauge models. An analogue of the plaquette formulation for the principal chiral model is so-called 
link representation \cite{link_spin,link_pt}. In this case one can construct a local dual theory 
for all $U(N)$ and $SU(N)$ principal chiral models \cite{spin_dual}. 
\item 
Dual representations based on 1) the character expansion of the Boltzmann 
weight and 2) the integration over link variables using Clebsch-Gordan expansion \cite{gauge_dual,dual4d}. 
This approach is not very useful in the context of principal chiral models as the summation over group indices 
cannot be performed locally. But in the case of LGT, due to the gauge invariance the summation over group (colour) 
indices can be done and this results in the local formulation in terms of invariant $6j$ symbols. This dual form 
can be studied using Monte-Carlo simulations \cite{MCdual,dual_comp1}. 
\item 
In the strong coupling limit the $SU(N)$ LGT can be mapped onto 
monomer-dimer-closed baryon loop model \cite{karsch}. 
\item 
More recent interesting approaches are developed in \cite{nlink_action} and in \cite{su2_dual}. 
\end{itemize} 
Important application of dual formulations concerns gauge models at finite baryon chemical potential. 
In some cases the sign problem can be explicitly solved in frameworks of the dual approach. This is 
the case, {\it e.g.} for the massless two-dimensional $U(1)$ LGT \cite{2dqed}. Also, the sign problem 
can be fully eliminated in the $SU(3)$ spin model in the complex magnetic field in the flux representation 
for the partition function \cite{spin_flux1,spin_flux2}. This model is an effective Polyakov loop model 
which can be calculated from the QCD partition function at strong coupling and large quark masses. 

In this contribution we present another approach to the duality transformations for $U(N)$ spin and gauge models. 
Our approach is based on the Taylor expansion of the Boltzmann weight and an exact integration over original 
gauge or spin degrees of freedom. Integrals to be calculated had been studied in the large $N$ limit in 
the end of the seventies \cite{weingarten}. Important functions which appear after such integration are called 
now the Weingarten functions, and their theory have been well developed during last decade \cite{collins1,collins2,novaes}. 
In Sect.~\ref{sec-1} we introduce our notations and present main results about group integrals which we need here. 
In Sects.~\ref{sec-2}-\ref{sec-5} we list main applications related to the construction of dual formulations 
in several cases. The main results and perspectives are outlined in Sect.~\ref{sec:discussion}.

\section{U(N) group integrals}\label{sec-1} 

\subsection{Notations and conventions}\label{notations}

We work on a $d$-dimensional hypercubic lattice $\Lambda = L^d$
with $L$ - a linear extension and a unit lattice spacing. 
$\vec{x}\equiv x=(x_1,...,x_d)$, $x_i\in [0,L-1]$ denote the sites of the lattice, 
$l=(\vec{x},\mu)$ is the lattice link in the $\mu$-direction and $p=(\vec{x},\mu<\nu)$ is 
the plaquette in the $(\mu,\nu)$-plane. $e_{\mu}$ is a unit vector in the direction $\mu$. 
We impose the periodic boundary conditions (BC) in all directions. 
Let $G=U(N)$; $U(x), U_l=U_{\mu}(x)\in G$, and $dU$ denotes the Haar measure
on $G$. ${\rm Tr}U$ will denote the fundamental character of $G$. We treat
models with a local interaction whose partition functions can be written as 
\begin{eqnarray} 
\label{PF_spindef}
&&Z_{\Lambda}(\beta,h_r,h_i;N)  \equiv  Z_{\rm {spin}} \ = \ 
\int \ \prod_x \ dU(x)  \\ 
&&\times \ \exp \left [ \beta_s \ \sum_{x,\mu} \ {\rm Re}{\rm Tr}U(x){\rm Tr}U^{\dagger}(x+e_{\mu}) + 
\sum_x \left ( h_r {\rm {Tr}}U(x) + h_i {\rm {Tr}}U^{\dagger}(x) \right ) \right ] \  \nonumber 
\end{eqnarray}
in case of $U(N)$ spin models and 
\begin{equation} 
Z_{\Lambda}(\beta;N)  \equiv  Z_{\rm {gauge}} \ = \ 
\int \ \prod_l \ dU_l  \ 
\exp \left [ \beta_g \ \sum_p \ {\rm Re}{\rm Tr}U(p) \ \right ] 
\label{PF_gaugedef} 
\end{equation}
in case of $U(N)$ LGT, where the plaquette matrix reads 
\begin{equation}
U_p \ = \ U_{\mu} (x)U_{\nu} (x+\mu)U_{\mu}^{\dagger}(x+\nu)U_{\nu}^{\dagger}(x) \ . 
\label{plaq}
\end{equation}
$U(N)$ spin model can be considered as an effective model for the Polyakov loop in the finite-temperature 
$U(N)$ LGT with $N_f$ flavours of massive staggered fermions. One has 
\begin{equation} 
h_r \ = \ \frac{1}{2} \ \sum_{f=1}^{N_f} \ h_f \ e^{\mu_f} \ , \  
h_i \ = \ \frac{1}{2} \ \sum_{f=1}^{N_f} \ h_f \ e^{- \mu_f} \ ,  \ 
h_f \ = \ (\cosh m_f)^{-1} \ . 
\label{hri}
\end{equation}
The relation between couplings reads $\beta_s\sim\beta_g^{N_t}$, with $N_t$ temporal extent of the lattice. 

\subsection{Group integrals and Weingarten function}\label{weingarten}

The basic integral which we need below is of the form 
\begin{equation} 
{\cal I}_N(r,s) \ = \  \int  dU  \prod_{k=1}^{r} U^{i_k j_k} \prod_{n=1}^{s} U^{m_n l_n *} 
\ = \ \delta_{r,s} \ {\cal I}_N(r)\ . 
\label{unintg_def}
\end{equation}
Its large-$N$ asymptotic behaviour was investigated in \cite{weingarten} and  
calculated in \cite{collins1} for $r \leq N$ and extended to $r > N$ 
in \cite{collins2} (a simple proof can be found in Ref.~\cite{novaes})  
\begin{equation} 
{\cal I}_N(r) \ = \ \sum_{\tau,\sigma\in S_r}  Wg^N(\tau^{-1}\sigma) \prod_{k=1}^{r}
\delta_{i_k,m_{\sigma(k)}} \delta_{j_k,l_{\tau(k)}} \ .
\label{basic_intg}
\end{equation}
$S_r$ is a group of permutations of $r$ elements and  
$Wg^N(\sigma)$ is the Weingarten function which depends only on the length of the cycles 
of a permutation $\sigma$. Its explicit form is given by 
\begin{equation} 
Wg^N(\sigma) = \frac{1}{(r!)^2} \ \sum_{\lambda} \ \frac{d^2(\lambda)}{s_{\lambda}(1)} \ \chi_{\lambda}(\sigma) \ , 
\label{Wgn_def}
\end{equation}
where $d(\lambda), \chi_{\lambda}(\sigma)$ are the dimension and the character of 
the irreducible representation $\lambda$ of $S_r$. The irreducible representations $\lambda$ are enumerated 
by partitions $\lambda=(\lambda_1,\lambda_2,\cdots,\lambda_{l(\lambda)})$ of $r$, where 
$l(\lambda)$ is the length of the partition and $\lambda_1\geq\lambda_2\cdots\lambda_{l(\lambda)}> 0$. 
The sum in (\ref{Wgn_def}) is taken over all $\lambda$ such that $l(\lambda)\leq N$. 
$s_{\lambda}(X)$, $X = (x_1,x_2,\cdots,x_N)$, is the Schur function ($s_{\lambda}(1)$ is the dimension 
of $U(N)$ representation). 

In the context of $U(N)$ spin model we need an integral of the form 
\begin{equation} 
Q_N(r) \ = \ \int dU \ \left [ {\rm Tr} U \ {\rm Tr} U^* \right ]^r \  . 
\label{trace_intg}
\end{equation}
Obviously, $Q_1(r)=1$. For arbitrary $N$ the compact and simple result can be obtained with the help 
of basic formula (\ref{basic_intg}). Namely, expanding the traces as sums over diagonal elements one gets 
\begin{eqnarray}
&&Q_N(r) \ = \ \sum_{i_1i_2\cdots i_r=1}^N \ \sum_{j_1j_2\cdots j_r=1}^N \ 
\int dU \ \prod_{k=1}^r \ U_{i_ki_k} U_{j_kj_k}^*   
\ = \ r! \ \sum_{i_1i_2\cdots i_r=1}^N \ \sum_{\sigma\in S_r}  Wg^N(\sigma) \prod_{k=1}^{r} \ 
\delta_{i_k,i_{\sigma(k)}} \nonumber  \\ 
&&=r! \ \sum_{\sigma\in S_r}  Wg^N(\sigma) \ P_{\sigma}(I) \ = \  \sum_{\lambda} \ d^2(\lambda) \ , 
\label{trace_intg1}
\end{eqnarray}
where $P_{\sigma}(I)=N^{|\sigma|}$ is the power sum symmetric function of a unit argument 
and we used the relation 
\begin{equation} 
s_{\lambda}(X) \ = \ \sum_{\tau\in S_r} \ \chi_{\lambda}(\tau) \ P_{\tau}(X) \ . 
\label{schur_pwrsum}
\end{equation} 
A few remarks are in order here. 
The constraint $r=s$ appearing in (\ref{unintg_def}) is essentialy abelian one. One solves the constraint 
by introducing genuine dual variables, like in $U(1)$ model. No other constraints are generated. 
For $U(N)$, in any dimension, summation over group (matrix) indices is factorized 
in every lattice site and can be done locally. Therefore, the dual theory is a theory with only local interaction. 
Last property holds also in the presence of fermions.

\section{One link integrals}\label{sec-2} 

The simplest one-link integral which gives an exact solution of two-dimensional pure  
$U(N)$ LGT 
\begin{equation} 
Z \ = \ \int \ dU \ \exp \left [ \beta \ {\rm Re}{\rm Tr}U \ \right ] \ = \ 
{\rm det} \ I_{i-j}(\beta) \ , \ i,j=1,\cdots,N \ 
\label{linkint_1} 
\end{equation}
can be easily computed. Expanding exponential in the Taylor series and using (\ref{trace_intg1}) one finds 
\begin{equation} 
Z \ = \ \sum_{r=0}^{\infty} \ \left ( \frac{\beta}{2} \right )^{2r} \ 
\frac{1}{(r!)^2} \ Q_N(r) \ . 
\label{linkint_2} 
\end{equation}
Another one-link integral appears in the strong coupling limit $\beta_g=0$ of $U(N)$ LGT with $N_f$ flavours 
of staggered fermions 
\begin{equation}
Z_0 \ = \ \int dU \ \prod_{f=1}^{N_f} \ 
\exp \left[ \eta_{x, \mu} \left( \bar{\psi}_f^{i}(x) U^{ij} \psi_f^{j}(x+\mu) -
\bar{\psi}_f^i(x+\mu) U^{\dagger,ij}\psi_f^j(x) \right) \right] \ . 
\label{linkint_ferm_def}
\end{equation} 
With the help of (\ref{basic_intg}) it is strightforward to obtain 
\begin{equation}
Z_0 \ = \ \sum_{r=0}^{N N_f} \ \sum_{\tau \in S_r} \frac{(-1)^{\tau}}{r!} \ Wg^N(\tau) \ 
\sum_{f_k, v_k = 1}^{N_f} \ \prod_{k=1}^{r} \ \sigma_{f_k, v_k}(x) \ \sigma_{v_k, f_{\tau(k)}}(x+\mu) \ , 
\label{linkint_ferm}
\end{equation} 
where $\sigma_{f, v}(x)$ is colourless meson field with $f,v$ flavour indices 
$\sigma_{f, v}(x) = \sum_{i = 1}^{N}  \bar{\psi}_f^i(x) \ \psi_v^i(x)$. 
The notation $(-1)^{\tau}$ means (-1) if the permutation $\tau$ is odd and +1 if it is even.
For one flavour one recovers the well-known result \cite{rossi}.

\section{Polyakov loop model}\label{sec-3}

In this Section we present our results for the Polyakov loop model defined in (\ref{PF_spindef}). 
Expanding all exponentials in the Taylor series one again encounters the integral of the form (\ref{trace_intg}). 
Computing all integrals with the help of Eq.(\ref{trace_intg1}) and making the change of summation variables 
suggested in \cite{spin_flux1} for $SU(3)$ spin model we write down the partition function in the form 
\begin{eqnarray}
&&Z_{\rm {spin}} \ = \  \sum_{r(l)=-\infty}^{\infty} \ \sum_{p(l)=0}^{\infty} \ \sum_{t(x)=0}^{\infty} \ 
\prod_l \left [ \left ( \frac{\beta}{2} \right )^{| r(l) | + 2p(l)} \ 
\frac{1}{(p(l)+| r(l) |)!p(l)!} \right ]  \nonumber  \\
&&\prod_x \left [ \frac{(h_r h_i)^{t(x)+\frac{1}{2}|r(x)|}}{t(x)!( t(x)+ |r(x)| )!} \ Q_N(s(x)) \right ] \ , 
\label{Zspin1}
\end{eqnarray} 
\begin{equation}
s(x) \ = \ \sum_{i=1}^{2d} \ \left ( p(l_i) + \frac{1}{2} \ | r(l_i) |   \right ) + t(x) + 
\frac{1}{2} \ |r(x)| \ , \  r(x) = \sum_{n=1}^d \ \left ( r_{\mu}(x) - r_{\mu}(x-\mu)   \right ) \ . 
\label{spin_sx}
\end{equation} 
where $l_i$ are $2d$ links attached to a site $x$. The Boltzmann weight is strictly positive if $h_r h_i \geq 0$. 
When external fields are vanishing $h_r=h_i=0$, only configurations $t(x)=r(x)=0$ contribute to the partition function. 
The constraint $r(x)=0$ can be solved in any number of dimensions by introducing dual variables. For example, 
in two-dimensional model we find the following representation for the partition function on the dual lattice 
\begin{equation}
Z_{\rm {spin}} \ = \  \sum_{r(x)=-\infty}^{\infty} \ \sum_{p(l)=0}^{\infty} \ 
\prod_l \left [ \left ( \frac{\beta}{2} \right )^{| r(x)-r(x+\mu) | + 2p(l)} \ 
\frac{1}{(p(l)+| r(x)-r(x+\mu) |)!p(l)!} \right ]  \ \prod_p \ Q_N(s(p)) \ , 
\label{Zspin1_dual}
\end{equation}
\begin{equation}
s(p) \ = \ \sum_{i=1}^{4} \ \left ( p(l_i) + \frac{1}{2} \ | r(l_i) |   \right ) \ .  
\label{spin_sx_dual}
\end{equation} 
Here, $l_i$ are 4 links forming dual plaquette $p$ and $r(l_1)=r(x)-r(x+\mu)$ and so on. 
For the $U(1)$ model, $Q_1(s)=1$, we recover the conventional dual form of the two-dimensional $XY$ spin model 
\begin{equation}
Z_{\rm {spin}} \ = \  \sum_{r(x)=-\infty}^{\infty} \ 
\prod_l \ I_{ r(x)-r(x+\mu)}(\beta)  \ ,  \  I_r(x)- {\rm {the \ modified \ Bessel \ function}} \ . 
\label{Zspin_U1}
\end{equation}

\section{Pure gauge models}\label{sec-4}

Here we turn our attention to pure gauge models. In order to use the integration method of Sect.~\ref{sec-1} 
we expand the integrand of (\ref{PF_gaugedef}) in the Taylor series and express the traces of the plaquette matrices 
as sums over group indices. The integration over link variables leads to a complicated set of Kronecker deltas 
on each link of the lattice. The main observation is that on every link this set of deltas is divided into two 
subsets. Each subset can be identified with one of lattice sites a given link belongs to. Thus, all summations 
over group indices are factorized in every lattice site in any dimension. This is a direct consequence of the local symmetry. 
In each site the combination of the sets of deltas, which come from all 
links containing given site, defines a site permutation $\gamma(x)$ on the set
of all group indices corresponding to this site. This site permutation $\gamma(x)$ is
a function of the link permutations $\sigma_l,\tau_l$.
The lengthy calculations of the corresponding sums will be presented in a separate publication. Here 
we present our results for three- and four-dimensional models. In $3d$ the constraint of Eq.(\ref{unintg_def}) $\delta_{r,s}$ 
takes a form $r(l)=r(p_1)+r(p_2)-r(p_3)-r(p_4)=0$, where $p_i$ are four plaquettes having link $l$ in common, and can be solved 
by introducing dual variables and placing them in the centers of original cubes. 
Then, the summation over group indices and duality transformations lead to the following representation  
on the dual lattice 
\begin{equation}
Z_{\rm {gauge}} =  \sum_{r(x)=-\infty}^{\infty} \ \sum_{k(l)=0}^{\infty} \  \sum_{\{ \tau_p,\sigma_p \}} \ 
\prod_l \ \frac{\beta^{2k(l)+|r(x)-r(x+\mu)|}}{k(l)!(k(l)+|r(x)-r(x+\mu)|)!} \ 
\prod_p Wg^N(\tau^{-1}_p\sigma_p) \  \prod_{c} \ P_{\gamma(c)}(1) \ . 
\label{Z_gauge}
\end{equation}
Here, $\sigma_p,\tau_p$ are elements of a permutation group $S_P$, $P=\sum_{l\in p}(k(l) + |r(x)-r(x+\mu)|/2)$. 
$\prod_c$ runs over all elementary cubes of the dual lattice and the symmetric function $P_{\gamma(c)}(1) = N^{|\gamma(c)|}$. 
$|\gamma(c)|$ is the number of cycles in combined permutations: 
$\gamma(c) \in S_C$,  $C=\sum_{l\in c}(2 k(l) + |r(x)-r(x+\mu)|)$. 
Similar to the spin models one recovers the conventional dual form for $U(1)$ LGT
\begin{equation}
Z = \sum_{r(x)=-\infty}^{\infty} \ \prod_l \  I_{r(x)-r(x+\mu)}(\beta) \ . 
\end{equation} 
Dual representation can be simplified by using orthogonality relation for the characters 
\begin{equation} 
\sum_{\omega\in S_r} \ \chi_{\mu}(\omega\tau) \ \chi_{\lambda}(\omega \sigma) \ = \ 
r! \ \delta_{\mu,\lambda} \ \frac{\chi_{\lambda}(\tau^{-1}\sigma)}{d(\lambda)} \ . 
\end{equation}
One then finds
\begin{equation}
Z_{\rm {gauge}} = \sum_{r(x)=-\infty}^{\infty} \ \sum_{k(l)=0}^{\infty} \ \sum_{\{ \omega_p \}} \ \prod_l \ 
\frac{\beta^{2k(l)+|r(x)-r(x+\mu)|}}{k(l)!(k(l)+|r(x)-r(x+\mu)|)!} \ \prod_{c} \ 
\left ( \ \prod_{p\in c} \ \sum_{\sigma_p} \ B_p \ \right ) \  N^{|\gamma(c)|} \ , 
\label{Z_gauge1}
\end{equation}
where we introduced notation 
\begin{equation}
B_p = \sum_{\lambda} \ \left ( \ \frac{d(\lambda)}{r!} \ \right )^{3/2} \ \frac{1}{(s_{\lambda}(1))^{1/2}} \ 
\chi_{\lambda}(\omega_p \sigma_p) \ . 
\end{equation}

The above consideration can be directly generalized to the four-dimensional theory. The only important 
difference is that in four dimension the solution of the constraint for the original plaquette variables $r(p)$
is given by the dual link variables. Therefore, the dual of the partition function can be written as 
\begin{equation}
Z_{\rm {gauge}} =  \sum_{r(l)=-\infty}^{\infty} \ \sum_{k(p)=0}^{\infty} \  \sum_{\{ \tau_c,\sigma_c \}} \ 
\prod_p \ \frac{\beta^{2k(p)+|r(p)|}}{k(p)!(k(p)+| r(p) |)!} \ 
\prod_c Wg^N(\tau^{-1}_c\sigma_c) \  \prod_{h} \ P_{\gamma(h)}(1) \ , 
\label{Z_gauge_4d}
\end{equation}
where 
\begin{equation}
r(p) \ = \ r(l_1)+r(l_2)-r(l_3)-r(l_4) \ ,
\end{equation}
and $l_i$ are four links forming a given oriented plaquette $p$. $\prod_c$ runs over all elementary cubes of the dual lattice. 
$\sigma_c,\tau_c$ are elements of a permutation group $S_C$, $C=\sum_{p\in c}(k(p) + |r(p)|/2)$. 
$\prod_h$ runs over all hypercubes of the dual lattice and the symmetric function $P_{\gamma(h)}(1) = N^{|\gamma(h)|}$. 
$|\gamma(h)|$ is the number of cycles in combined permutations which belong to the permutation group  
$S_H$,  $H=\sum_{p\in h}(2 k(p) + |r(p)|)$. The generalization of Eq.(\ref{Z_gauge1}) to four dimension is also strightforward.

\section{Two-dimensional U(N) QCD}\label{sec-5}

Finally, we consider two-dimensional $U(N)$ LGT with one flavour of the massive staggered 
fermions. The presence of the dynamical fermions does not destroy the main observation 
of the previous section, namely all summations over group indices are factorized around 
lattice sites and can be performed as before. The resulting permutation group becomes even 
more complicated as it includes now the link occupation numbers. All details of the integration 
and summation will be given in a separate work. We present here the final result for 
the partition function. Expressed in terms of the plaquette and link occupation numbers 
the partition function reads 
\begin{eqnarray} 
&&Z = \sum_{r(p)=-\infty}^{\infty} \ \sum_{t(p)=0}^{\infty} \
\sum_{ k(l) =0}^{N} \ \sum_{ n(l)=0}^{N} \ \sum_{ s(x) =0}^{N} \ 
\sum_{\{\tau_l, \sigma_l\}\in S_X(l)} \ \prod_x \ m^{s(x)} \ N^{|\gamma(x)|}
\nonumber  \\    
&&\prod_p \frac{(\beta/2)^{2t(p)+|r(p)|}}{t_p!(t_p + |r(p)|)!}  
\prod_l \ \left [ \frac{\eta_{\nu}(x)}{2} \right ]^{k(l)+n(l)} \ 
e^{-\mu (k(l)-n(l))} \ Wg^N(\tau^{-1}_l\sigma_l)  \nonumber  \\   
&&\times \ ( {\rm {constraints}} ) \ \times \ ( {\rm {sign \ factor}} ) \ ,
\label{2d_un}
\end{eqnarray} 
where $\eta_{\nu}(x)$ is staggered sign factor, $\mu$ - chemical potential. 
Permutation group $S_X(l)$ is fixed by
\begin{equation} 
X(l) = t(p)+t(p^{\prime}) + \frac{1}{2} \ \left ( |r(p)|+r(p)+|r(p^{\prime})| - 
r(p^{\prime}) \right ) + k(l) \ .
\end{equation}
Factor $N^{|\gamma(x)|}$ arises after summation over all group indices,
$|\gamma(x)|$ is the number of cycles in combined permutations $\sigma_l,\tau_l$, $l\in x$. 
Allowed configurations of monomers, plaquette and link occupation numbers are constrained 
by the integration over gauge fields on every link 
\begin{equation} 
\prod_l \ \delta \left ( r(p) - r(p^{\prime}) + k(l) - n(l) \right )  \ , \ \ 
p,p^{\prime} \  {\rm {have \ common \ link}} \ l , 
\label{gauge_constr}
\end{equation}
and by the integration over fermion fields in every site
\begin{eqnarray}
&&\prod_x \ \delta \left ( s(x)+k(x) -N \right ) \ \delta \left ( s(x)+n(x) -N \right ) \ , 
\nonumber  \\ 
&&k(x) = \sum_{\nu=1}^2 \ [ k_{\nu}(x) + n_{\nu}(x-\nu) ] \ , \ 
n(x) =  \sum_{\nu=1}^2 \ [ n_{\nu}(x) + k_{\nu}(x-\nu) ] \ .
\label{ferm_constr}
\end{eqnarray}
Sign factor appears due to the integration over fermions and is non-trivial only in the presence of 
closed fermion loops as allowed by the above constraints. It takes a form similar to   
the massless two-dimensional QED \cite{2dqed}, namely 
\begin{equation}
\prod_{\cal{L}} \ (-1)^{1+\frac{1}{2}|\cal{L}|} \ ,
\label{sign_2d}
\end{equation}
where $\cal{L}$ is a closed fermion loop and $|\cal{L}|$ is a length of a given loop. Detailed derivation 
of this formula as well as discussion of the general conditions for which the full Boltzmann weight is positive 
will be given elsewhere.

\section{Discussion}\label{sec:discussion} 

In this paper we presented a new approach to construction of the dual formulations 
of non-abelian models. The existing integration methods, explained in Sect.~\ref{sec-1}, allow  
to make such dual transformations for $U(N)$ lattice spin and gauge models. We have constructed 
dual forms of Polyakov loop spin models for arbitrary $U(N)$ model with and without external field. 
Also, we have outlined our main findings for pure $U(N)$ gauge theories and for two-dimensional 
$U(N)$ QCD. Details of our calculations will be reported in future publications. 

Some obvious extensions of this work can be done. For example, it should not be difficult to 
generalize the present approach for gauge models with Wilson fermions. More important and interesting 
problem concerns extension of this method to $SU(N)$ LGT. The recent paper \cite{zuber} can be helpful 
in this direction. 

As for the applications of our dual formulations we think it can be interesting in, at least, two aspects. 
Since the asymptotic expansions and bounds on the Weingarten function are known it might give a new direction 
in the investigation of LGT at large values of $N$. It can be useful in the studying of the confinement problem, as well. 
It remains to be seen if the dual formulation can help in solving the sign problem in QCD, at least in 
two-dimensional theories.



\end{document}